\documentclass[dvips]{article}
\usepackage{icrctc07,epsfig,url}
\def\agt{\raise0.3ex\hbox{$\;>$\kern-0.75em\raise-1.1ex\hbox{$\sim\;$}}}
\def\d{{\rm d}}
\newcommand{\be}{\begin{equation}}
\newcommand{\ee}{\end{equation}}
\newcommand{\bea}{\begin{eqnarray}}
\newcommand{\eea}{\end{eqnarray}}

\title{High Energy Neutrinos with a Mediterranean Neutrino
Telescope}
\shorttitle{High Energy Neutrinos with a Mediterranean Neutrino
Telescope}
\authors{E. Borriello$^{1,2}$, A. Cuoco$^3$, G. Mangano$^1$, G. Miele$^{1,2}$, S. Pastor$^2$, O. Pisanti$^1$, P.~D.~Serpico$^4$}
\shortauthors{E. Borriello et al}
\afiliations{$^1$Dipartimento di Scienze Fisiche, Universit\`a di
Napoli ``Federico II" and INFN Sezione di Napoli, Com\-plesso
Universitario di Monte S. Angelo, Via Cintia, Napoli, 80126, Italy \\
$^2$Instituto de F\'{\i}sica Corpuscular (CSIC-Universitat de
Val\`{e}ncia), Ed. Institutos de Investigaci\'{o}n, Apdo. 22085,
E-46071 Valencia, Spain \\
$^3$Institut for Fysik og Astronomi, Aarhus Universitet Ny
Munkegade,
Bygn. 1520 8000 Aarhus Denmark \\
$^4$Center for Particle Astrophysics, Fermi National Accelerator
Laboratory, Batavia, IL 60510-0500, USA}
\email{pisanti@na.infn.it}

\abstract{The high energy neutrino detection by a km$^3$ Neutrino
Telescope placed in the Mediterranean sea provides a unique tool
to both determine the diffuse astrophysical neutrino flux and the
neutrino-nucleon cross section in the extreme kinematical region,
which could unveil the presence of new physics. Here is performed
a brief analysis of possible NEMO site performances.}

\begin{document}
\maketitle

\section{Introduction}
Neutrinos are one of the main components of the cosmic radiation in
the high energy regime.  Although their fluxes are uncertain and
depend on the production mechanism, their detection can provide
information on the sources and origin of the high energy cosmic
rays.

From the experimental point of view the detection perspectives are
stimulated by the Neutrino Telescopes (NT's) constructed, like
\verb"Baikal" \cite{Balkanov:1999rf} and \verb"AMANDA"
\cite{Andres:1999hm}, or under construction like \verb"IceCube"
\cite{Ahrens:2003ix} under the ice and \verb"ANTARES"
\cite{antares} in the deep water of the Mediterranean sea. Here,
also the experiments \verb"NESTOR" \cite{nestor} and \verb"NEMO"
\cite{nemo} are in the R\&D phase and, together with
\verb"ANTARES", in the future could lead to the construction of a
km$^3$ telescope as pursued by the \verb"KM3NeT" project
\cite{km3net}.

Although NT's were originally thought as $\nu_{\mu}$ detectors,
their capability as $\nu_{\tau}$ detectors has become a hot topic in
view of the fact that flavor neutrino oscillations lead to nearly
equal astrophysical fluxes for the three neutrino flavors. Despite
the different behavior of the produced tau leptons with respect to
muons in terms of energy loss and decay length, both $\nu_\mu$ and
$\nu_{\tau}$ event detection rates are sensitive to the matter
distribution near the NT area. In principle, the elevation profile
of the Earth surface around the detector may be relevant. In Ref.
\cite{Miele}, some of the present authors calculated the aperture of
the Pierre Auger Observatory \cite{Abraham} for Earth-skimming UHE
$\nu_{\tau}$'s, by using the Digital Elevation Map (DEM) of the site
(GTOPO30) \cite{GTOPO30}. In Ref. \cite{Cuoco} the DEM's of the
under-water Earth surface, provided by the Global Relief Data survey
(ETOPO2) \cite{ETOPO2}, was used to estimate the effective aperture
for $\nu_\tau$ and $\nu_\mu$ detection of a km$^3$ NT in the
Mediterranean sea placed at any of the three locations proposed by
the \verb"ANTARES", \verb"NEMO" and \verb"NESTOR" collaborations. In
the present paper we further develop the approach of Ref.
\cite{Cuoco} to evaluate the performances of a Mediterranean NT in
the simultaneous determination of the neutrino flux and the
$\nu$-Nucleus cross section in extreme kinematical regions (which
may probe new physics, see e.g. \cite{AlvarezMuniz:2001mk}). Since
the three different proposed sites for the under-water km$^3$
telescope show event rate differences of the order of 20\%, for the
sake of brevity we report the results our analysis for the
\verb"NEMO" site only, which presents intermediate performances.

\section{Formalism and results}

Following the formalism developed in \cite{Miele,Cuoco} we define
the km$^3$ NT {\it fiducial} volume as that bounded by the six
lateral surfaces $\Sigma_a$ (the index $a$=D, U, S, N, W, and E
labels each surface through its orientation: Down, Up, South, North,
West, and East), and indicate with $\Omega_a \equiv (\theta_a,
\phi_a)$ the generic direction of a track entering the surface
$\Sigma_a$ (see Figure 4 of Ref. \cite{Cuoco} for notations). We
introduce all relevant quantities with reference to $\nu_\tau$
events, the case of $\nu_\mu$ being completely analogous.

Let $\d \Phi_\nu/(\d E_\nu \, \d\Omega_a)$ be the differential
flux of UHE $\nu_\tau + \bar{\nu}_\tau$. The number per unit time
of $\tau$ leptons emerging from the Earth surface and entering the
NT through $\Sigma_a$ with energy $E_\tau$ is given by
\bea\label{e:1}
\left( \frac{ \d N_\tau}{\d t} \right)_a = && \!\!\!\!\!\!\!\!\!\!
\int \d\Omega_a \int \d S_a \int \d E_\nu \, \frac{\d \Phi_\nu
(E_\nu, \Omega_a)}{\d E_\nu\, \d \Omega_a} \nonumber \\
\int \d E_\tau && \!\!\!\!\!\!\!\!\! \cos\left(\theta_a\right)
k_a^\tau (E_\nu,E_\tau;\vec{r}_a,\Omega_a).
\eea

The kernel $k_a^\tau(E_\nu\,,E_\tau\,;\vec{r}_a,\Omega_a)$
represents the pro\-bability that an incoming $\nu_\tau$ crossing
the Earth, with energy $E_\nu$ and direction $\Omega_a$, produces a
$\tau$-lepton which enters the NT fiducial volume th\-rough the
lateral surface $\d S_a$ at the position $\vec{r}_a$ with energy
$E_\tau$. For an isotropic flux and an exposure time $T$, the total
number of $\tau$ leptons (and similarly for muons) crossing the NT
is \bea\label{e:kernel1} N_\tau = T~ \sum_a \int \d\Omega_a \, \int
\d S_a \int \d E_\nu
\int \d E_\tau \nonumber \\
\left( \frac{1}{4\pi}\, \frac{\d \Phi_\nu(E_\nu)}{\d E_\nu}
\right) \cos \left( \theta_a \right) \, k_a^\tau
(E_\nu\,,E_\tau\,; \vec{r}_a,\Omega_a).
\eea

Although the exact dependence of Eq. (\ref{e:kernel1}) on the
neutrino flux and the neutrino-nucleon charged current cross section
$\sigma_{CC}^{\nu N}$ may be quite complicated, basic physical
considerations show that even a rough binning of the events for
energy loss and arrival direction may be used to obtain information
on both these quantities (see e.g.
\cite{Hooper:2002yq,Hussain:2006wg}). In particular, in the
following we shall consider the sum of the $\mu$ and $\tau$
contributions as the experimental observable, namely the energy
deposited in the detector and not the energy and/or the nature of
the charged lepton crossing the NT. In fact, only for a minor
fraction of the detected events the nature of the charged lepton can
be reliably established.

According to Ref.s \cite{Aramo,Dutta2}, the differential energy
loss of the $\tau$ leptons per unit of length in an underwater NT
can be simply taken as $\d E_\tau/\d \lambda=- \beta_\tau \,
E_\tau \varrho_w$, with $\beta_\tau = 0.71 \times 10^{-6}$ cm$^2$
g$^{-1}$ and $\varrho_w$ denoting the water density. Analogously,
for muons one just needs to replace $\beta_\tau$ with the
corresponding value $\beta_\mu=0.58 \times 10^{-5}$ cm$^2$
g$^{-1}$. Assuming that the lepton energy loss in the NT by e.m.
interactions, $\Delta E_l$, is just a small fraction of its energy
at the entrance, $E_l$, we simply obtain $\Delta E_l =
\lambda(\vec{r}_a,\Omega_a)\, \beta_l \, E_l\, \varrho_w$, where
$\lambda(\vec{r}_a,\Omega_a)$ is the length crossed in the NT by
the lepton whose track is defined the geometrical quantities
$\vec{r}_a,\Omega_a$.

Using these relations one can derive the spectrum of leptons
detected in the NT as a function of their deposited energy,
$\Delta E$, and their arrival direction, $\Omega\equiv
(\theta,\phi)$, measured in the zenith-azimuth reference frame
\bea\label{e:kernel3}
\frac{\d^2 N}{\d (\Delta E) \d \Omega} = && \!\!\!\!\!\!\!\!\!\! T
\sum_{\alpha=\mu,\tau} \sum_a \int \d S_a \int \d E_\nu \nonumber \\
\frac{1}{4\pi} && \!\!\!\!\!\!\!\!\!\!\! \frac{\d
\Phi_\nu(E_\nu)}{\d E_\nu}~ \frac{\cos \left( \theta_{a}\right)\,
k_a^\alpha} {\lambda ( \vec{r}_a, \Omega_a )
\beta_\alpha\, \varrho_w}.
\eea
By denoting with $X_i$ a given bin in energy loss, and with $Y_j$
the one for the zenithal angle, we can integrate the expression
(\ref{e:kernel3}) to get the number of expected events in $X_i
\times Y_j$,
\bea\label{e:kernel4}
N_{ij} = && \!\!\!\!\!\!\!\!\!\! T \sum_{\alpha=\mu,\tau} \sum_a
\int_{X_i} \d (\Delta E) \int_{Y_j} \d \Omega \\
\int && \!\!\!\!\!\!\!\!\!\!\!\! \d S_a \int \d E_\nu
\frac{1}{4\pi}\, \frac{\d \Phi_\nu(E_\nu)}{\d E_\nu}~ \frac{\cos
\left( \theta_{a}\right)\, k_a^\alpha} {\lambda ( \vec{r}_a,
\Omega_a ) \beta_\alpha\, \varrho_w}. \nonumber
\eea

To take into account the underwater surface profile one can
numerically compute the above integral as described in Ref.
\cite{Miele}: by using the available DEM of the area near the
\verb"NEMO" site, one can isotropically generate a large number of
oriented tracks which cross the \verb"NEMO" fiducial volume (see
Figure 4 of Ref. \cite{Cuoco}) and sample the above integrand. This
technique allows also to account for the radial density profile of
the Earth (we use the formula reported in \cite{Gandhi96}).

In order to study the sensitivity to both neutrino flux and
$\sigma_{CC}^{\nu N}$ it is necessary to parameterize their
standard expressions and the possible departures from them. In
particular, we parameterize the flux as $\d \Phi_\nu/\d E_\nu\, \d
\Omega_a = C\cdot 1.3\cdot 10^{-8}\,
\left(E_\nu/\mbox{GeV}\right)^{-2\, D}$ GeV$^{-1}$ cm$^{-2}$
s$^{-1}$ sr$^{-1}$, which gives a standard Waxman-Bahcall flux
\cite{Waxman} for $C = D = 1$. For the neutrino-nucleon cross
section we use:
\begin{eqnarray}
\frac{\sigma_{CC}^{\nu N}}{0.344 \, \textrm{nb}} = \left\{
\begin{array}{ll}
 \!\!\! \left(\frac{E_\nu}{E_1}\right)^{0.492\,A} & \!\!\!\!\!\!
E_\nu \leq E_2\\
 \!\!\! \left(\frac{E_2}{E_1}\right)^{0.492\,A} \left(\frac{E_\nu}{E_2}\right)^{0.492\,B}
 & \!\!\!\!\!\! E_{\nu} > E_2
\end{array}
\right. \nonumber
\end{eqnarray}

where $E_1=10^{5.5}\,$GeV is the energy below which the
atmospheric flux is expected to dominate (so we consider only the
region $E>E_1$) and $E_2=10^{6.0}\,$GeV. In the low-energy bin
this cross-section matches the standard expression \cite{Gandhi98}
for $A=1$. A value of $B$ significantly larger than 1 may be
associated with new physics. Note that the factor $C$ only enters
via the product $CT$ as a normalization and can be fixed to $C=1$,
considering instead the exposure time $T$ as the effective
variable.

\begin{figure}
\begin{center}
\includegraphics[width=.4\textwidth]{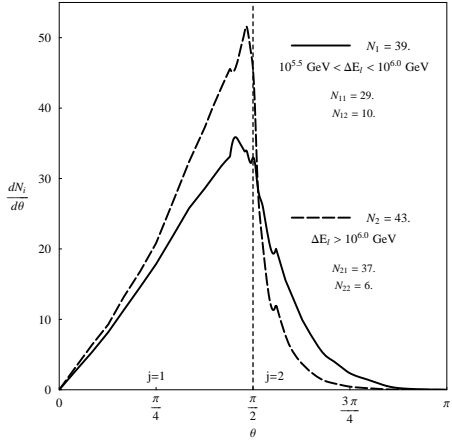}
\end{center}
\caption{Angular distributions of ($\mu+\tau$) events collected in
five years from a km$^3$ NT placed at the NEMO site (see text).}
\label{f:th_dist}
\end{figure}

For illustrative purposes, in Figure \ref{f:th_dist} we report the
event angular distribution, for a km$^3$ NT placed at the
\verb"NEMO" site in five years of operations. The solid and dashed
lines correspond to events whose energy loss in the detector
belongs to the intervals $10^{5.5}$-$10^6$ GeV or $>10^6$ GeV,
respectively. The predictions are obtained for standard flux and
cross section ($A=B=C=1$). In the plot are also reported the
number of events $N_{ij}$ (see Eq. (\ref{e:kernel4})) when we
consider $i=1,2$ for the previous two energy bins and $j=1,2$ when
the zenith arrival direction is between 0$^\circ$ and 90$^\circ$
or 90$^\circ$ and 180$^\circ$.

\begin{figure}
\begin{center}
\includegraphics[width=.4\textwidth]{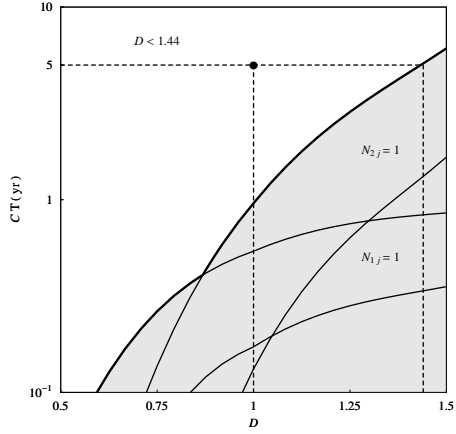}
\end{center}
\caption{$(C T,D)$ region corresponding to the observation of at
least one event in each bin (for standard cross section).}
\label{f:CD_range}
\end{figure}

Clearly, for a very steep flux power-law index $D$, the number of
events decreases. We shall require that at least one event falls
in each bin, in $T$ years of running, in the case of standard
cross-section; this rough criterion constrains the parameter range
that one experiment is able to explore to the brighter region of
Fig. \ref{f:CD_range}, corresponding to the intersection of the
regions where $N_{ij}\geq 1$, for all $i,j$.

\begin{figure}
\begin{center}
\includegraphics[width=.35\textwidth]{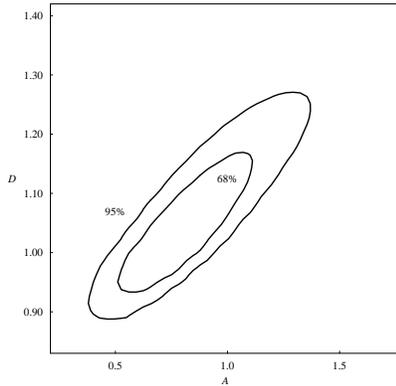}
\end{center}
\caption{Marginalized contour levels in the $(A,D)$ plane (for
$A=B$) (see text for details).} \label{f:contour}
\end{figure}

As a preliminary result, in Fig. \ref{f:contour} we show the
constraints (contours at the 68 \% and 95 \% CL) which can be
obtained on the physical parameters $A$ and $D$ after the
marginalization over $C$ is made. Here we are assuming $B=A$, so
that the plot represents the capability of the telescope to
disentangle the energy dependence of the flux from the energy
dependence of the cross-section (in the toy model where both are
described by a single parameter). We performed a multi-Poisson
likelihood analysis \cite{Baker}, in which the likelihood
function, $L=\exp(-\chi^2/2)$, is defined using the following
expression for the $\chi^2$ ($N_{ij}^0$ being the event numbers of
the reference model): \be \chi^2=2\sum_{ij}\left[
(N_{ij}-N_{ij}^0)+N_{ij}^0\ln(N_{ij}^0/N_{ij}) \right]. \ee

\section{Conclusions}
We have performed an analysis of the capability of a km$^3$ NT in
the Mediterranean to disentangle the high energy neutrino flux and
neutrino-nucleon cross section in an unexplored kinematical
region. Our statistical analysis exploits the dependence of
observables on energy and arrival direction (under the hypothesis
of an isotropic diffuse flux). Using a simplified toy model to
parameterize fluxes and cross-sections, preliminary results
confirm that this approach is very promising, and could
potentially detect hints of new physics. Of course the real
feasibility of such measurements will depend crucially on the size
of the neutrino flux which fixes the time required to reach a
reasonable statistics. A complete account of this research will be
reported in a forthcoming publication.
\\
\\
\noindent {\bf Acknowledgements:} P.S. acknowledges support by the
US Department of Energy and by NASA grant NAG5-10842. G. Miele
acknowledges support by Generalitat Valenciana (ref. AINV/2007/080
CSIC).


\begin{thebibliography}{30}

\bibitem{Balkanov:1999rf}
  V.~A.~Balkanov {\it et al.}  [Baikal Collaboration],
  Phys.\ Atom.\ Nucl.\  {\bf 63}, 951 (2000)
  [Yad.\ Fiz.\  {\bf 63N6}, 1027 (2000)]
  [arXiv:astro-ph/0001151].

\bibitem{Andres:1999hm}
  E.~Andres {\it et al.},
  Astropart.\ Phys.\  {\bf 13}, 1 (2000)
  [arXiv:astro-ph/9906203].

\bibitem{Ahrens:2003ix}
  J.~Ahrens {\it et al.}  [IceCube Collaboration],
  Astropart.\ Phys.\  {\bf 20}, 507 (2004)
  [arXiv:astro-ph/0305196].

\bibitem{antares}
  M.~Spurio  [ANTARES Collaboration],
  arXiv:hep-ph/0611032.

\bibitem{nestor}
  G.~Aggouras {\it et al.}  [NESTOR Collaboration],
  Nucl.\ Instrum.\ Meth.\  A {\bf 567}, 452 (2006).

\bibitem{nemo}
  P.~Piattelli  [NEMO Collaboration],
  Nucl.\ Phys.\ Proc.\ Suppl.\  {\bf 165}, 172 (2007).

\bibitem{km3net}
  U.~F.~Katz,
  Nucl.\ Instrum.\ Meth.\  A {\bf 567}, 457 (2006)
  [arXiv:astro-ph/0606068].

\bibitem{Miele}
  G.~Miele, S.~Pastor and O.~Pisanti,
  Phys.\ Lett.\  B {\bf 634}, 137 (2006)
  [arXiv:astro-ph/0508038].

\bibitem{Abraham}
  J.~Abraham {\it et al.}  [Pierre Auger Collaboration],
  Nucl.\ Instrum.\ Meth.\  A {\bf 523}, 50 (2004).

\bibitem{GTOPO30}
U.S. Geological Survey's Center for Earth Resources Observation and
Science (EROS), 1996, \url{http://asterweb.jpl.nasa.gov}

\bibitem{Cuoco}
  A.~Cuoco  {\it et al.},
  JCAP {\bf 0702}, 007 (2007)
  [arXiv:astro-ph/0609241].


\bibitem{ETOPO2}
National Geophysical Data Center, 2001,
   \url{http://www.ngdc.noaa.gov/mgg/fliers/01mgg04.html}

\bibitem{AlvarezMuniz:2001mk}
  J.~Alvarez-Muniz, F.~Halzen, T.~Han and D.~Hooper,
  Phys.\ Rev.\ Lett.\  {\bf 88}, 021301 (2002)
  [arXiv:hep-ph/0107057].

\bibitem{Hooper:2002yq}
  D.~Hooper,
  Phys.\ Rev.\  D {\bf 65}, 097303 (2002)
  [arXiv:hep-ph/0203239].

\bibitem{Hussain:2006wg}
  S.~Hussain, D.~Marfatia, D.~W.~McKay and D.~Seckel,
  Phys.\ Rev.\ Lett.\  {\bf 97}, 161101 (2006)
  [arXiv:hep-ph/0606246].

\bibitem{Aramo}
  C.~Aramo  {\it et al.},
  Astropart.\ Phys.\  {\bf 23}, 65 (2005)
  [arXiv:astro-ph/0407638].

\bibitem{Dutta2}
  S.~I.~Dutta, Y.~Huang and M.~H.~Reno,
  Phys.\ Rev.\  D {\bf 72}, 013005 (2005)
  [arXiv:hep-ph/0504208].

\bibitem{Gandhi96}
  R.~Gandhi, C.~Quigg, M.~H.~Reno and I.~Sarcevic,
  Astropart.\ Phys.\  {\bf 5}, 81 (1996)
  [arXiv:hep-ph/9512364].


\bibitem{Waxman}
  E.~Waxman and J.~N.~Bahcall,
  Phys.\ Rev.\  D {\bf 59}, 023002 (1999)
  [arXiv:hep-ph/9807282].

\bibitem{Gandhi98}
  R.~Gandhi, C.~Quigg, M.~H.~Reno and I.~Sarcevic,
  Phys.\ Rev.\  D {\bf 58}, 093009 (1998)
  [arXiv:hep-ph/9807264].

\bibitem{Baker}
  S.~Baker and R.~D.~Cousins,
  Nucl.\ Instrum.\ Meth.\  A {\bf 221}, 437 (1984).
   \end{thebibliography}
\end{document}